\documentclass{elsart}

\usepackage{amssymb}
\usepackage{graphicx}
\usepackage{lineno}
\usepackage{placeins}
\usepackage[numbers]{natbib}

\begin{document}
%\linenumbers

\begin{frontmatter}



\title{Response of microchannel plates to single particles and to electromagnetic showers}


%\author[label2]{M.Barnyakov},
\author[label0]{L.~Brianza}, 
\author[label1]{F.~Cavallari}, 
\author[label1]{D.~Del Re}, 
\author[label1]{S.~Gelli},
\author[label0]{A.~Ghezzi},
\author[label0]{C.~Gotti}, 
\author[label0]{P.~Govoni}, 
\author[label1]{C.~Jorda},
\author[label0]{A.~Martelli},
\author[label0]{B.~Marzocchi}, 
\author[label1]{P.~Meridiani},
\author[label1]{G.~Organtini},
\author[label1]{R.~Paramatti},
\author[label0]{S.~Pigazzini}, 
\author[label1]{S.~Rahatlou},
\author[label1]{C.~Rovelli},
\author[label1]{F.~Santanastasio},
\author[label0]{T.~Tabarelli~de~Fatis \corauthref{cor1}},
\author[label0]{N.~Trevisani}

%\address[label2]{Buderk Ins
\address[label0]{Universit\`a di Milano Bicocca and INFN, Sezione di Milano-Bicocca, \\
                          Piazza della Scienza 3, I-20126, Milano, Italy}
%\address[label2]{INFN, Sezione di Roma1,\\ 
%                          P.le A.~Moro 1, I-00044 Rome, Italy}
\address[label1]{Sapienza Universit\`a di Roma and INFN, Sezione di Roma1,\\ 
                          P.le A.~Moro 1, I-00044 Rome, Italy}

\corauth[cor1]{Corresponding author: tommaso.tabarelli@mib.infn.it}

\begin{abstract}

We report on the response of microchannel plates (MCPs) to single
relativistic particles and to electromagnetic showers. Particle
detection by means of secondary emission of electrons at the MCP
surface has long been proposed and is used extensively in ion
time-of-flight mass spectrometers. What has not been investigated in
depth is their use to detect the ionizing component of showers. The
time resolution of MCPs exceeds anything that has been previously used
in calorimeters and, if exploited effectively, could aid in the event
reconstruction at high luminosity colliders. Several prototypes of
photodetectors with the amplification stage based on MCPs were 
exposed to cosmic rays and to 491~MeV electrons at the INFN-LNF
Beam-Test Facility.  The time resolution and the efficiency of the 
MCPs are measured as a function of the particle multiplicity, and the 
results used to model the response to high-energy showers.

\end{abstract}

\end{frontmatter}

% main text
\section{Introduction}
\label{intro} 

With the packet structure of beams at hadron colliders, high
luminosities are achieved at the cost of an increased number of 
concurrent collisions per beam crossing at the experiment collision
point. At the Large Hadron Collider (LHC), there are typically
20$\div$30 overlapping interactions per beam crossing, spread over a
length of about 5~cm root mean square (RMS) along the beam axis. Event
reconstruction exploits the association of individual particles to an
interaction vertex: tracks or energy deposits inconsistent with the
vertex of interest are filtered, or statistically subtracted. This
approach becomes strained at the High-Luminosity LHC (HL-LHC)
\textendash~and, in prospect, at future colliders~\textendash, where
about 140$\div$200 collisions per beam crossing are anticipated. With
peak vertex densities above 1~mm$^{-1}$, tracks from  nearby vertices
could be merged into a fake, high-energy event vertex. 
More importantly, even at moderate vertex
densities, the random overlap of energy deposits from neutral
particles \textendash~mainly photons~\textendash, which cannot be
tied to any vertex, deteriorates the performance of calorimeters in
terms of energy measurement and particle identification, as particles
appear to be less isolated. A precise measurements of the time of the
energy deposits and of each collision vertex has been advocated as a
means to mitigate these effects~\cite{White:2014oga}. Due to the
length of the packets, collision vertices at the HL-LHC have an RMS
time spread of about 200~ps. A time resolution of about 20~ps would
therefore reduce the `effective multiplicity' of concurrent collisions
by a factor 10, down to a level comparable to the LHC. This resolution
is one order of magnitude better than at current LHC
experiments~\cite{Abelev:2014ffa}, \cite{Aad:2014gfa},
\cite{DelRe:2015}.

In this work, we characterize the response of microchannel plates
(MCPs) \cite{Wiza:1979} to single relativistic particles and to the
ionizing component of electromagnetic showers. Due to their superior
time response, a layer of MCPs embedded in a calorimeter, or in a
preshower compartment of it, could be exploited to provide a precise 
measurement of the photon time. In addition, even for moderate
efficiencies to minimum ionizing particles, the time of each vertex
could be reconstructed from the time of energy deposits associated to
charged tracks, owing to the large track multiplicity at hadron
colliders. This detector would therefore enable, at once, time
separation of vertices in spatial overlap and assignment of the
neutral energy to the proper vertex. A preshower would factorize the
quest for precision timing from the technological choice of the full
calorimeter in future experiments, or could be added in front of
existing calorimeters in an upgrade program of current detectors.

The use of secondary emission of electrons at the MCP surface to
sample the ionizing component of showers was pioneered in
1990~\cite{Derevshchikov:1990ej}. The MCP response to relativistic
particles was also investigated in the '90s, and detection
efficiencies of around 70\%, with time resolution of 70~ps were
achieved~\cite{Bondila:2002sy}.  Recent technological progress of the
Large Area Picosecond Photodetector Collaboration~\cite{Adams:2013nva}
may result in a reduction of the cost production for MCPs and is
spurring a renewed interest in this detection techniques. A set of
measurements - similar in scope to our work - has been recently
reported in~\cite{Ronzhin:2014lqa}, where the response of
photodetectors based on MCP multipliers was tested in proton and
positron beams. Time resolution of order 20$\div$30~ps at shower
maximum were obtained. The contribution to the detector response from  
secondary emission at the MCP surface was indirectly inferred by
estimating the contribution from Cherenkov emission in the optical
window of the photodetector at different window thicknesses. 

At variance with that work, we directly measure and characterize the
secondary emission from the MCP surface in MCP-based photomultipliers   
(PMT-MCP), by applying a retarding bias to the photocathode, in order
to inhibit avalanche formation from electrons emitted at the
photocathode. We refer to this setup as to an 'ionization-MCP'
(i-MCP). Measurements are also reported for the PMT-MCP operation
mode,  where Cherenkov emission from the photodetector window was
exploited. The potential advantage of an i-MCP consists in the
elimination of the photocathode, resulting in a more robust design and
in a potentially improved radiation tolerance, since radiation damage
mostly affects the photocathode response~\cite{Lehmann:2014aqa}.

After the description of the detectors and of the measurement setup
(Sec.~\ref{detector}), we present results on the response to single
particles and to showers at different depths in units of radiation
lengths (Sec.~\ref{time} and \ref{performance}). A response model for
the MCPs is developed along with the discussion of the results, and
then used to anticipate the performance of a preshower detector
(Sec.~\ref{extra}). Ways to further improve the response of i-MCPs are
also indicated. 

\section{Detector description and operation modes}
\label{detector}

Our measurements are carried out with PMT-MCPs developed at BINP
(No\-vo\-si\-birsk), in collaboration with the Ekran FEP manufacturer. Full 
characterization of these photodetectors is reported in
Ref.~\cite{Barnyakov:2007zza}. Four MCP-PMTs were made available for
this study\footnote{Courtesy of M.~Barnyakov}. All the devices have an
18~mm diameter and 1.2~mm thick optical window, made of borosilicate
glass coated with multialkali photocathode, which provides a peak
quantum efficiency of about 15\% at 500~nm. The amplification stage
consists of two stacked layers of about 0.4~mm thick MCP wafers made
of lead glass, with channel diameters ranging between 7 and 10~$\mu$m,
and channel pitch ranging between 10 and 12~$\mu$m, depending on the
device. The channels in the first and second MCP layers have a bias
angle to the photodetector axis of 5$^o$ and 13$^o$, respectively. The
photocathode is separated from the input stage of the MCP by a gap of
0.2~mm, while the gap from the MCP output to the anode is of 0.4~mm. 

In addition, we use also one Photonis-XP85012 PMT-MCP, comprising an
optical window 53$\times$53~mm$^2$ wide, and an amplification stage
made of two lead glass MCP layers - each 1~mm thick - with 25~$\mu$m
diameter channels. The optical window, 3~mm thick, is coated with
bialkali photocathode providing a peak quantum efficiency of about
22\% at 380~nm. The anode readout plane is segmented in 64 pads. In
our measurements, this granularity was not exploited, and a common
signal (of positive polarity) at the output of the second MCP was read
out. This configuration is not optimized for time measurements, as the
capacitance associated with the wide readout area impacts on the time
response. Therefore, data from these measurements are not exploited to
qualify the time performance of Photonis-XP85012, but only to study
the secondary emission from the MCPs. Noteworthy, the geometry of the
MCP wafers is different from Ekran FEP devices, but the ratio of the
channel diameter to the wafer thickness (known as `aspect ratio') is
exactly 1:40 in both cases. The bias angle is also similar. 

A voltage divider is used to provide about 90\% of the voltage drop
through the MCP layers. Two alternate configurations are used to
characterize the MCP response. In the `PMT-MCP mode', 10\% of the
voltage drop is equally shared between the photocathode-to-MCP and
MCP-to-anode gaps. In this configuration, the response of the detector
is driven by avalanches in the MCPs triggered by photoelectrons
following Cherenkov emission in the optical window. The mean number of 
photoelectrons for relativistic particles at normal incidence is
estimated to $\mu \simeq 3$ in the Ekran FEP PMTs and $\mu \simeq
15$$\div$20 in the Photonis-XP85012 PMT. In the `i-MCP mode', a
retarding bias is instead applied to the photocathode-to-MCP gap, to
prevent photoelectrons from reaching the MCP surface and triggering an
avalanche. In this configuration, the response of the detector is
uniquely determined by  secondary emission of electrons from the MCP
layers when crossed by ionizing particles. 

\section{Time measurements with cosmic-ray muons}
\label{time}

The time response of the Ekran FEP MCPs to single particles was
studied with cosmic-ray muons. The cosmic-ray stand consists of a  
stack of three MCP detectors aligned to the vertical. The upper and lower
detectors are Ekran FEP MCP photomultipliers operated in PMT-MCP mode
to provide the trigger and a reference for efficiency and time
measurements, upon the passage of a cosmic ray. The third MCP
detector, interposed between the other two, is used either in PMT-MCP
or in i-MCP mode. In this setup, the rate of muons within the
acceptance of the trigger is about 1~h$^{-1}$. In an alternate
configuration, to increase the acceptance, only two MCP detectors are
stacked and used to form the trigger. 

Anode signals, with rise time of order 1~ns, are delayed by about
10~ns relative to the trigger, and sampled on a 50~$\Omega$ load with
a Tektronix DPO7254, providing an 8-bit digitization of the waveforms
at 20~GSample/s with an input bandwidth of 2.5~GHz. On each trigger,
%fired by the coincidence of the upper and lower PMT-MCPs, 
the waveforms of all the detectors are recorded for offline
analysis. The time information is extracted from interpolated
waveforms via constant fraction discrimination (CFD), with a threshold
set at 40\% of the maximum amplitude. A time-over-threshold algorithm
was also tested, and provided comparable performance. Signals are
retained in the analysis, if their amplitude is five times larger than
the RMS of the electronic noise. 

\begin{figure*}[htb]
\centering
\includegraphics[width=0.49\linewidth]{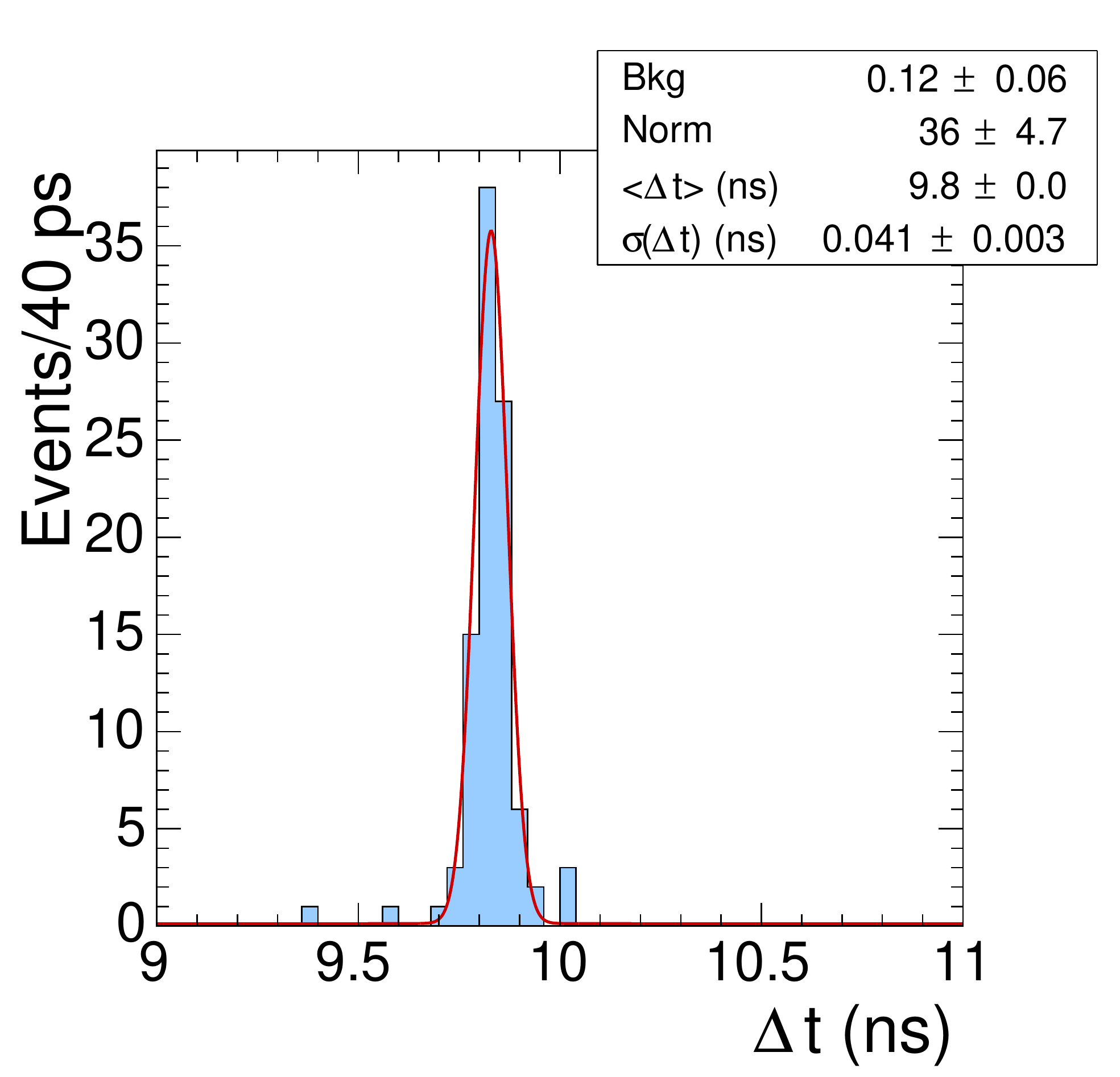}
\includegraphics[width=0.49\linewidth]{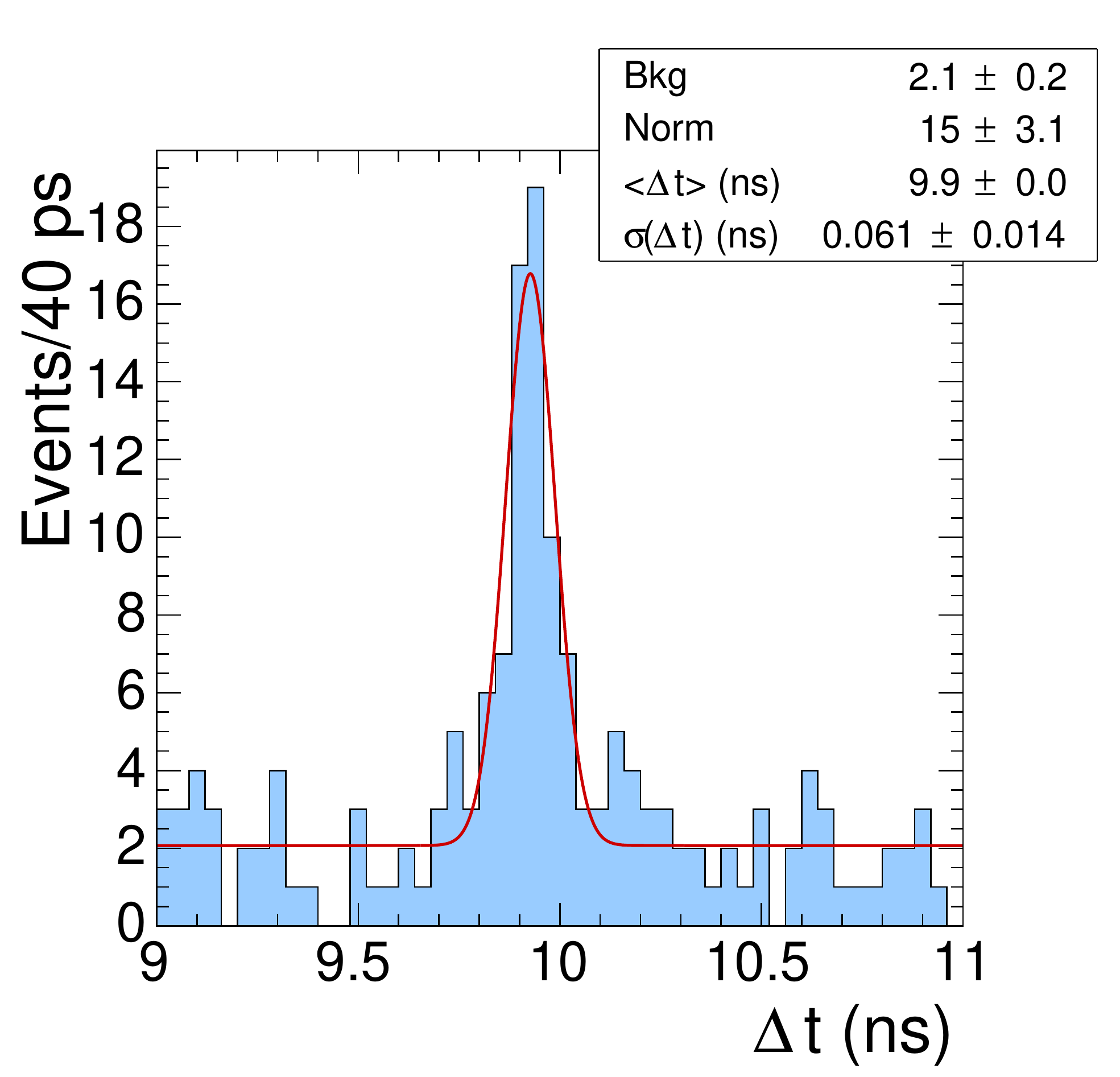}
\caption{Distribution of the time difference between signals due to
  cosmic ray muons through two PMT-MCPs, for signals generated by
  photoelectrons from Cherenkov emission in the optical window of the 
  two PMT-MCPs ({\it left}), and for signals generated by secondary
  emission of electrons at the MCP surface in one of the two PMT-MCPs 
  ({\it right}).}  
\label{timeCosm}
\end{figure*}

The distribution of the time difference between the MCP detector and
the PMT-MCP trigger signal is shown in Fig.\ref{timeCosm} for an
operating voltage of 2500~V. In the left panel, results are obtained
when both detectors are operated in PMT-MCP mode, while in the right  
panel one of the detectors is operated in i-MCP mode. A time spread of
40~ps is observed in the first case, corresponding to a time
resolution of 30~ps to relativistic charged particles in each 
detector. In the second configuration, the observed time spread is
60~ps, implying that the time resolution of the i-MCP detector to
single particles is about 50~ps. The larger rate of accidental
coincidences in i-MCP mode is due to different trigger configuration
and lower efficiency. While in PMT-MCP mode the detectors are fully
efficient to relativistic charged particles, in the i-MCP mode the
efficiency ranges between 10\% and 50\%, depending on the bias
voltage. Improved resolution and efficiency are expected for showers,
where the multiplicity of secondary particle crossing the MCPs is
higher. 

\section{Measurements with 491~MeV electrons} 
\label{performance}

\subsection{Setup at the LNF-BTF electron beam}
For further characterization of the response and measurement of the
efficiency to single particles and to showers, the MCP detectors were
exposed to an electron beam at the Beam Test Facility (BTF) of the INFN
Laboratori Nazionali di Frascati (Italy)~\cite{Ghigo:2003gy}. The beam provides
10~ns long electron pulses with tuneable energy (up to about 500~MeV),
repetition rate (up to 49~Hz) and intensity  (1$\div$10$^{10}$
particles/pulse). Our measurements were performed with 491~MeV
electrons and an intensity tuned to provide a mean multiplicity of
about one electron per pulse.

The MCP photomultipliers were mounted in a light-tight box with the
optical window towards the beam and orthogonal to the beam
direction. The first and last MCPs along the beam line were operated in 
PMT-MCP mode, to provide a reference event selection for efficiency 
measurements. A logic signal, synchronous with the beam gate, was used
to trigger waveform digitization of the anode signals, over 200~ns,
into a 12-bit 5~GSample/s switched capacitor digitizer
(CAEN-V1742). Delays were set to sample about 50~ns of baseline before
the signal pulse. Auxiliary beam counters upstream of the MCPs were
also readout into gated-ADCs for beam diagnostics and off-line
selection purposes. These include a 5~mm thick plastic scintillator
with 24$\times$24~mm$^2$ cross section and a scintillating fibre
hodoscope, covering an acceptance of $8\times8$~mm$^2$ with 1~mm pitch
in the two coordinates transverse to the beam. Further details on this
ancillary instrumentation are given in Ref.~\cite{Candelise:2014}.

Pulses consistent with a single electron entering the test setup,
identified from the pulse-height of the signal in the scintillator
beam counter, are retained in the offline analysis. Furthermore, the
fibre hodoscope is used to identify electrons within the geometric
acceptance of the MCPs. The selection is further refined by requesting
a pulse in the first PMT-MCP in the beam line, operated in PMT-MCP
mode. The charge and the time of each pulse in the downstream MCPs are
extracted upon integration and CFD discrimination of the individual
waveforms in a time window of 4~ns in coincidence with
the signal of the first PMT-MCP. Events are selected if the charge is
five times the RMS noise of the detector, measured from the
integration of a 4~ns window in the baseline region before the 
pulse. Empty beam pulses, i.e. with a signal in the scintillator
counter consistent with the pedestal, are used to estimate the rate
of accidental signals in the MCPs.

\subsection{Response to single particles}
\label{singles}

To measure the efficiency to single electrons, a coincidence is
required between the first and the last PMT-MCP along 
the beam line. Events in the MCP detector under study are accepted if
a signal with charge exceeding five times the RMS noise is found in
coincidence within 1~ns of the mean time of the reference PMT-MCP
signals. The raw efficiency result is corrected for random coincidences of
accidental signals falling within the same time window, which amount
to less than 0.1\% of the triggers.   

\begin{figure*}[htb]
\centering
\includegraphics[width=0.49\linewidth]{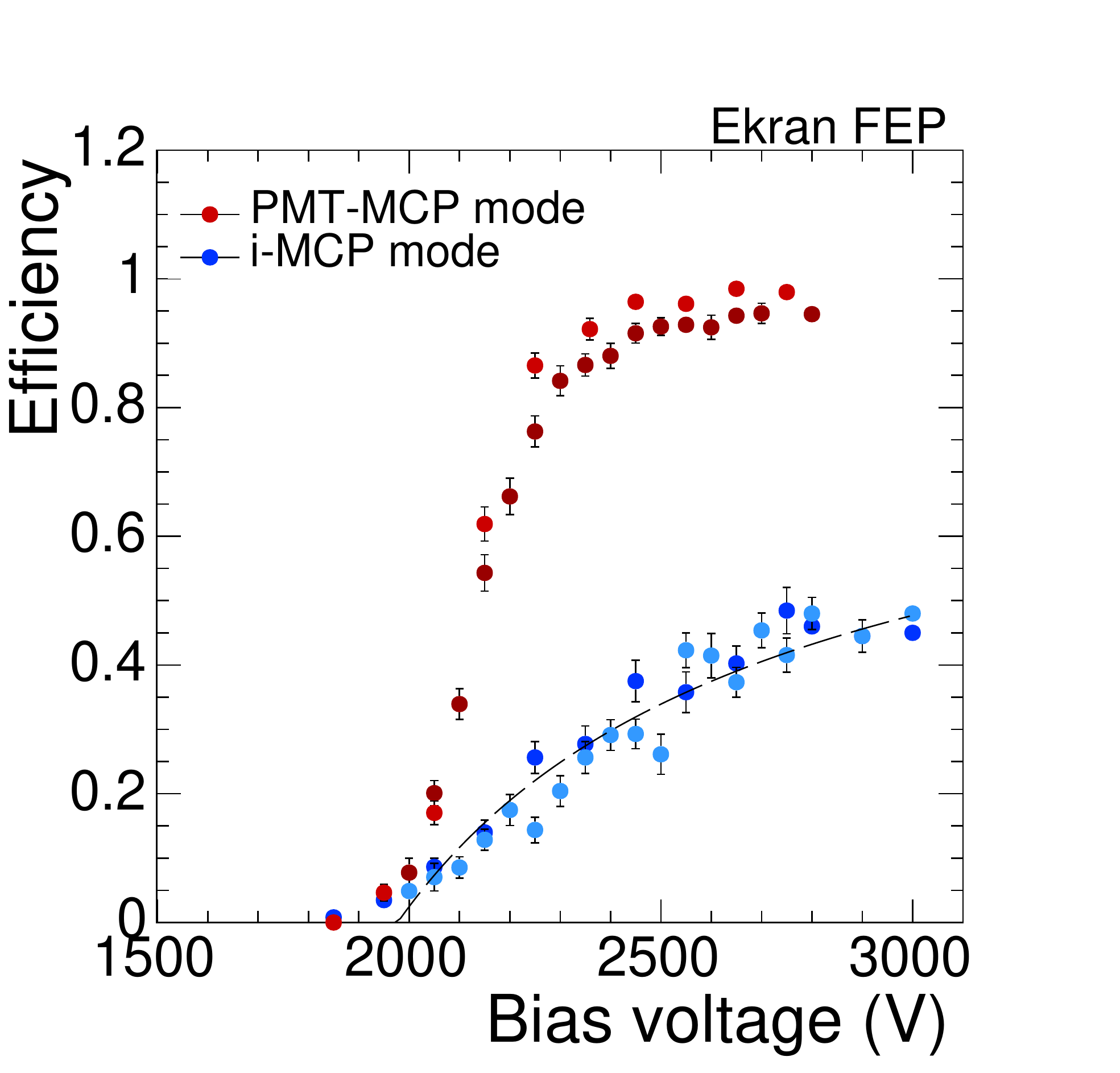}
\includegraphics[width=0.49\linewidth]{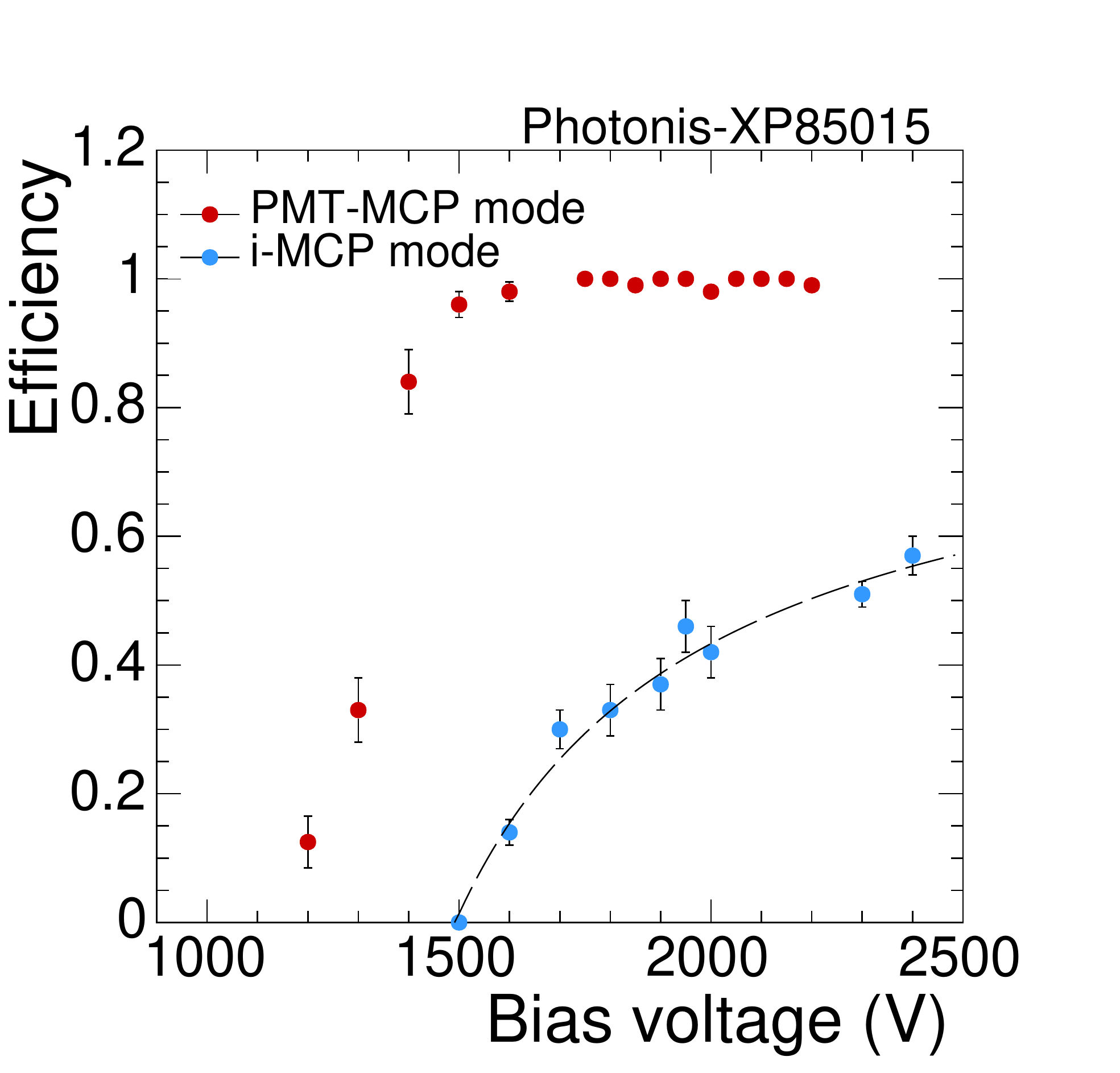}
\caption{Efficiency to 491~MeV electrons as a function of the bias
  voltage for MCP detectors with direct (red dots) and retarding
  bias (blue dots) between the photocathode and the first MCP
  layer. The curve through the points is a fit to data of a response
  model (see text for details). The left panel is for two Ekran FEP
  MCP detectors; the right panel for Photonis-XP85012.}
\label{fig1}
\end{figure*}

The efficiencies to 491~MeV electrons as a function of the bias voltage
are shown in Fig.~\ref{fig1}. Data in the left panel were collected
with Ekran FEP MCP photomultipliers operated in PMT-MCP mode (red 
dots) and in i-MCP mode (blue dots). Results in the right panel refer to 
Photonis-XP85012: due to limited beam time availability, they include
data collected at the cosmic muon test stand.

In PMT-MCP mode, with direct bias to the photocathode, a plateau
efficiency close to 100\% is achieved with all the devices. In this
configuration, secondary emission and amplification are operated by
separated elements of the detector: photoelectrons extracted from the
photocathode upon Cherenkov emission are amplified above detection
threshold by the MCPs. %From the properties of the optical window and
%of the photocathode, 
We estimate the mean number of photoelectrons to be $\mu \simeq 3$ and
15$\div$20 for the Ekran FEP and the Photonis-XP85012 photodetectors,
respectively. Data are consistent with a residual inefficiency of
$\exp(-\mu)$ at plateau, where the MCP gain is sufficient to supply single
photoelectron detection.

In i-MCP mode, a steady increase of the efficiency with the bias
voltage is observed above a threshold voltage not lower than the
threshold in PMT-MCP mode. A maximum efficiency in slight excess of 
50\% is achieved at the maximum high voltage operated during the
tests, but the curve is still not plateauing. In this configuration, the MCP
layers carry out the dual function of seeding the cascade process, via
the secondary electrons extracted from the MCP, and of providing 
signal amplification. The amplification process takes place over
different channel lengths, depending on the longitudinal position in
the channel where the secondary electron is extracted. Inefficiency
could arise either because of lack of secondary emission or because of
insufficient amplification in the cascade following the secondary
emission. The efficiency can therefore be written as:  
\begin{equation}
  \epsilon = s \left( 1 - \frac{L_{eff}}{L}\right),  
    \mbox{\hspace{4cm}}  (L_{eff} \leq L)
    \label{effeq}
\end{equation}
where $s$ (bound to $s\leq1$) indicates the probability of secondary
emission over the entire MPC length $L$, and $L_{eff}= L_{eff}(V)$ is
the minimum channel length that, at given voltage, provides sufficient
gain to overcome the detection threshold. In other words, up to gain
fluctuations, secondary emissions in the downstream section of the
channel of length $L_{eff}$ do not result in detectable signals;
secondary emissions from the complementary section of length 
$(L-L_{eff})$ generate detectable signals. Under the hypothesis that
the gain %over the entire channel 
has a power-law dependence on the
bias voltage, with power index proportional to %$L$, Eq.~(\ref{effeq})
to the channel length, 
Eq.~(\ref{effeq}) can be cast in the form:  
\begin{equation}
  \epsilon = s \left( 1 - \frac{1}{b \ln(V/V_{th})+1} \right), 
    \mbox{\hspace{2cm}}  (V \geq V_{th})
\label{eq2}
\end{equation}
where $s$, $V_{th}$ and $b$ are parameters to be extracted from data. At
the threshold voltage $V_{th}$, secondary electrons generated at the
upper surface of the MCP receive just the exact gain to become
detectable. Noteworthy, this threshold voltage corresponds to the
threshold for single photoelectron detection in PMT-MCP mode: at
$V<V_{th}$ the channel length that would be needed to supply
sufficient gain for a single photoelectron to be detected would 
exceed the physical length of the channel ($L_{eff}>L$). In agreement
with this, the threshold voltage is approximately the same  in both
operation modes with the Ekran FEP detectors (Fig.~\ref{fig1}-left),
for which the mean number of photoelectrons following Cherenkov
emission is just slightly above unity. In the Photonis-XP85012 
detector with $\mu \simeq 15$$\div$20, the efficiency threshold in
PMT-MCP mode occurs at an MPC gain that is about one order of
magnitude lower than in i-MCP mode, i.e. at a bias voltage about 200~V
lower than $V_{th}$ (Fig.~\ref{fig1}-right).

The dashed curves in Fig.~\ref{fig1} show the results of the fit to
the data of the model described by Eq.(\ref{eq2}). Consistency with
data is found for $s \simeq 1$, implying that there is at least one
secondary emission following the passage of an ionizing particle
through the two MCP wafers.  This means that the secondary emission
probability from a single wafer is at least 50\%. From the aspect
ratio and the bias angle of the MCPs, similar in all the devices under
study, we estimate that beam electrons at normal incidence on the MCP
detectors cross an MCP surface about 10 times as they go through one
wafer. We conclude that the secondary emission probability is of order
10\% per crossing of a channel surface by a single relativistic
particle.  

The dependence of the efficiency on the bias voltage is logarithmic,
and the efficiency gain as a function of the voltage too slow to be
practical. Moreover, the signal amplitude depends on where the 
secondary emission occurs, which may be sub-optimal for some
applications. Lines of investigations that we are pursuing to enhance
the response to single particles in i-MCPs include configurations with
multiple MCP stacks, larger bias angles, and wafers with enhanced
secondary emission.  These variations in assembly or in properties of
the wafers impact on the total amount of secondary emission and on the
channel gain. Based on our results, for example, we expect that a
stack of three MCP wafers could provide an efficiency of 70\% or more. 

While work is ongoing to refine the MCPs parameters, an efficiency of
50\% to single particles looks already promising for applications in
environments where the track multiplicity is high. This may be
sufficient, for example, to estimate the time of a collision vertex or
of electromagnetic showers.

\subsection{Response to electromagnetic showers}

Further insight in the response of i-MCP detectors is acquired from
the analysis of data collected with a set of absorbers of variable
thickness in front of the MCP detector under test. Similarly to the 
previous analysis, trigger counters and one PMT-MCP detector located
just upstream of the absorbers are used to identify beam pulses with
single electrons entering the test setup. No  requirements are instead
placed on MCPs downstream of the one under test, to prevent the
selection from biasing the sample with showers of multiplicity of
secondary particles higher than the average. Signals in the i-MCP
detector exceeding five times the RMS noise and in coincidence with
the PMT-MCP within 1~ns are searched for. The efficiency is measured
by counting the fraction of these events within the selected sample. 

\begin{figure*}[htb]
\centering
\includegraphics[width=0.8\linewidth]{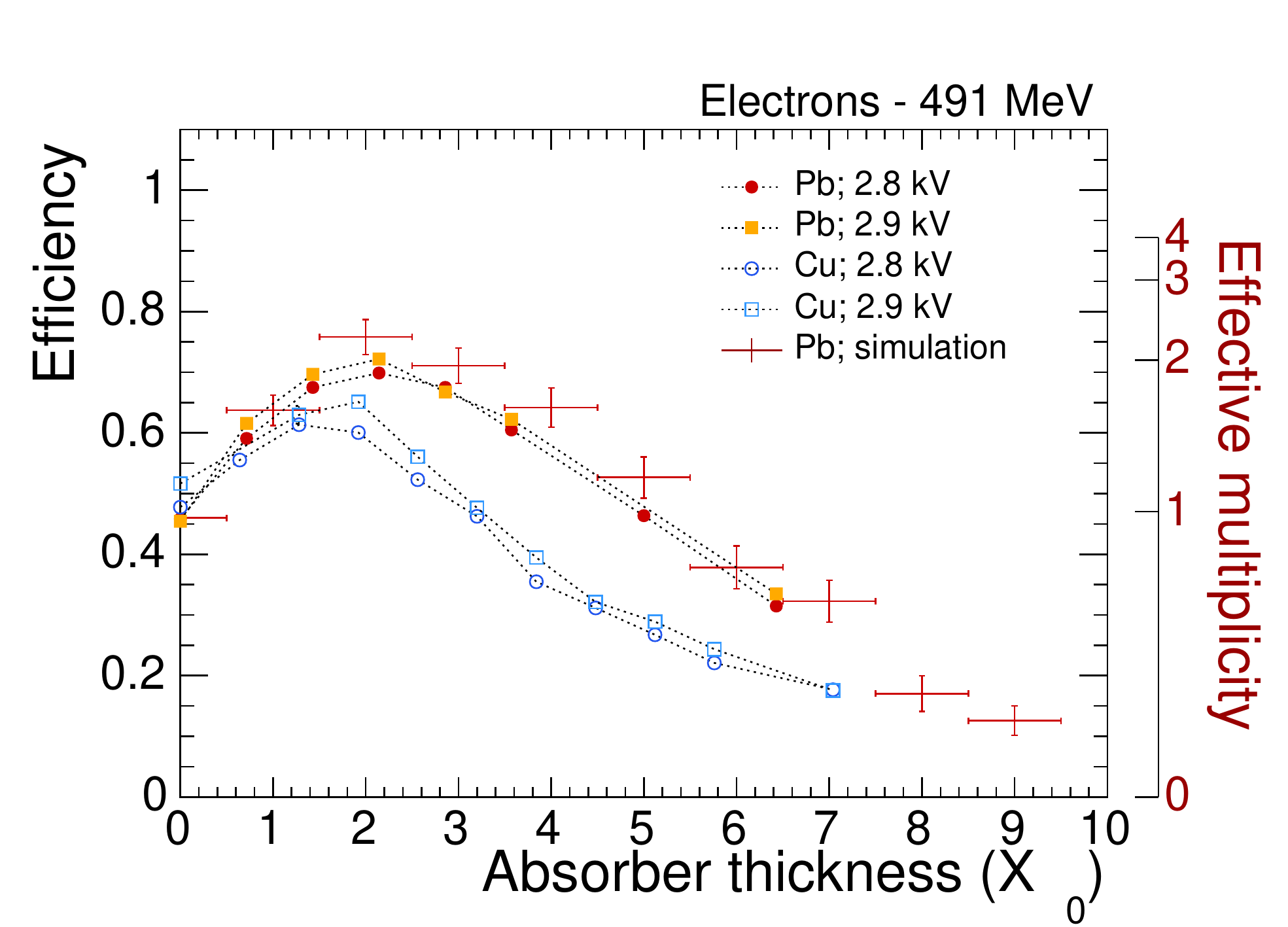}
\caption{Efficiency to 491 MeV single electrons in an i-MCP detector
  as a function of the thickness of lead (full symbols) and copper
  (open symbols) absorbers. A comparison to simulation for lead
  (crosses), and the effective multiplicity of particles crossing the
  MCPs are also shown (see text for details). Errors on the data
  points are smaller than the marker size. Lines are drawn to guide
  the eye.} 
\label{fig2}
\end{figure*}

The efficiency to 491~MeV electrons as a function of the absorber
thickness, in units of radiation lengths ($X_0$) for an operating
voltage of 2800 and 2900~V are shown in Fig.~\ref{fig2}. Results for
two different materials, copper and lead, are displayed. Since lead
has a lower critical energy, the multiplicity of secondary particles
is expected to be larger with lead absorbers than with copper ones.
The efficiency at zero thickness is consistent with the efficiency
measured in the analysis of the response to single electrons. As the
thickness of the lead absorber increases, the efficiency raises from
45\% to a maximum of about 70\% at a depth of about 2$X_0$. A
maximum at a shallower depth, and with an efficiency of about 65\% is
observed with copper absorbers. At larger thicknesses, the efficiency
decreases and eventually vanishes. This is understood as the effect of
the evolution of charged particle multiplicity within the shower for
491~MeV electrons. 

Data are also compared to a Monte Carlo (MC) simulation based on the
Geant4 package~\cite{Asai:2006qm,Allison:2006ve}.  Electrons of
491~MeV are fired on the absorber. The beam profile is tuned to match
the distribution measured at the hodoscope with the events selected by
the analysis. Secondary particles from the shower cascade are traced to
the MCP surface, and each charged particle with sufficient energy to
cross the full thickness of the wafers is assigned a probability
$\epsilon=45$\% to generate a detectable signal. The efficiency to
detect the shower is measured by the fraction of events in which at
least one secondary electron generated a signal. This is equivalent to
describe the MCP response with a binomial per-event probability:  
\begin{equation}
  \epsilon(n) = 1 - (1-\epsilon_1)^n,
  \label{binomial}
\end{equation}
where $\epsilon_1$ is the efficiency to single particles, and $n$ is
the multiplicity of charged particles crossing the MCP at a given
absorber depth. The result of the MC simulation, shown for the lead
absorber in Fig.\ref{fig2}, are in good agreement with data and
confirm our interpretation of the evolution of the response as a
function of the absorber depth. Equation~(\ref{binomial}) can be
inverted to express the particle multiplicity, as a function of the
efficiency. This is shown on the right axis of the plot in
Fig.~\ref{fig2}, for a single particle efficiency
$\epsilon_1=0.45$\%. According to this interpretation, the efficiency
at the shower maximum corresponds to an effective multiplicity of two
charged particles per 491~MeV electron.
 
\begin{figure*}[htb]
\centering
\includegraphics[width=0.49\linewidth]{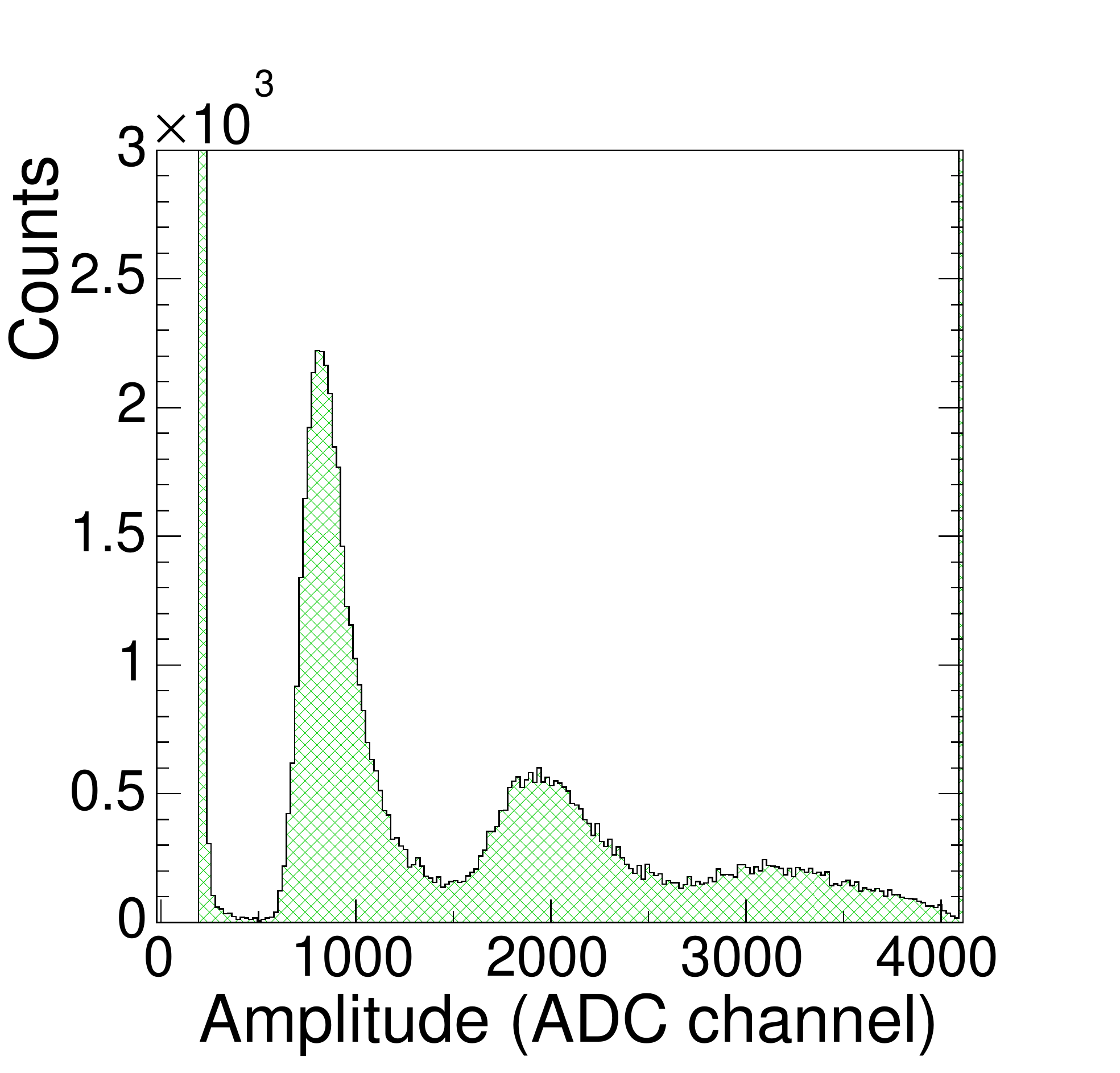}
\includegraphics[width=0.49\linewidth]{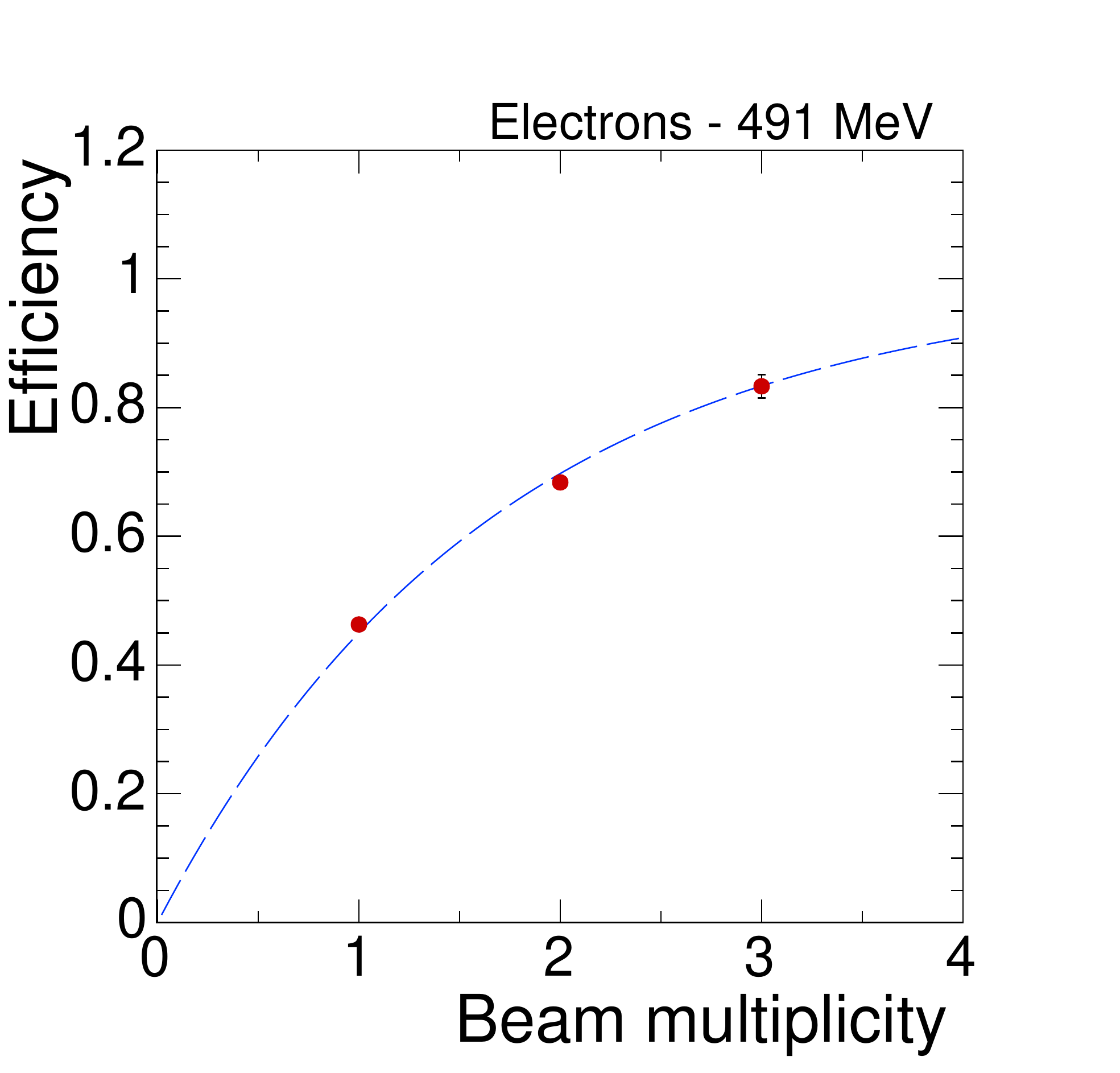}
\caption{Left: Pulse height spectrum of the scintillator counter with
  peaks corresponding to multiplicities of $n=1$, 2 and 3 electrons
  per pulse. Higher multiplicities saturated the ADC input of the test
  setup. Right: Efficiency of an Ekran FEP i-MPC at 2900 V as a function
  of the multiplicity of 491~MeV electrons. No absorber in front of
  the i-MPC is present. The curve shows a fit to data of the function
  of Eq.~(\ref{binomial}).} 
\label{fig4}
\end{figure*}

This interpretation is supported by a direct study of the MCP response
as a function of the beam multiplicity. The amplitude spectrum of the
scintillator counter, shown in the left panel of Fig.~\ref{fig4}, is
exploited to separate beam pulses with a multiplicity of $n=1$, 2 and
3 electrons. The analysis uses data collected with an Ekran FEP
photomultiplier with no absorber in front, operated in i-MCP mode at
2900~V. At each multiplicity, the efficiency $\epsilon(n)$ is measured
by counting the fraction of events with at least one signal higher
than five times the RMS noise, within a time window covering the beam
pulse duration. Spurious counts due to accidental coincidences are
subtracted with the same method as in the analyses previously
described. Results are shown in the right panel of Fig.~\ref{fig4},
where the efficiency is plotted as a function of the beam
multiplicity. Experimental data are well described by the function
of Eq.~(\ref{binomial}) with $\epsilon_1=45$\%, indicated by the
dashed line. A particle multiplicity in excess of four would bring the
efficiency to electromagnetic showers above 90\%. Any improvement in
the efficiency to single particles in future devices would directly
reflect in an improved response to showers.

\section{Extrapolation to photons of high energy}
\label{extra}

The simulation and the response model are extended to evaluate the
potential of photon timing at high energies with MCP detectors
embedded in a preshower. A detailed  design study is beyond the scope
of this document, but we focus on two aspects to help clarify some
performance requirements: on the one side, precision timing could be
fully exploited if achieved with sufficient efficiency, and on the
other, the impact on the energy resolution of the calorimeter system
should remain small. Both aspects are studied in dedicated
simulations. 

In the study of the efficiency, electrons and photons of 30 GeV are
simulated and propagated through the same MC simulation shown to match  
test beam results (see Sec.~\ref{performance}). These energies are
representative of physics processes relevant at the HL-LHC. We find that
one i-MCP detector with 45\% efficiency to single particles would be
sufficient to provide full efficiency to electrons and more than 
80\% efficiency for photons after 3$X_0$ of lead. The increase in
efficiency relative to direct measurements with 491~MeV electrons
should be ascribed to the higher multiplicity of secondary particles
in 30~GeV showers. The efficiency to photons would rise to above
90\%, with two sampling layers at 1$X_0$ and 3$X_0$, or alternatively
with a single layer, if the efficiency to single particles could be
increased to 70\%. The difference in efficiency between electrons and
photons reflects the conversion probability of photons in the absorber. 
At full MCP efficiency, a residual, irreducible inefficiency of
approximately 5\% to photons in a $3X_0$ preshower would still be
observed. As we commented earlier, work is ongoing to enhance the
single particle efficiency. 

To study the impact on the energy resolution, the simulation, used so
far only to count particles above the detection threshold, is
extended to include the analog response of i-MCPs. In i-MPC mode, the
gain of the detector depends on the position along the MCP channel
where the secondary emission occurs.  As a consequence, the amplitude 
spectrum observed with single particles at the test beam is broad
and approximately flat above the threshold amplitude up to a maximum
amplitude. This is implemented in simulation by randomly sampling a
uniform distribution for $\epsilon=45$\% of the particles crossing the
MCP detectors, randomly selected, and ascribing zero amplitude to the
remaining particles. Even if the single particle response is broad,
the mean response is still proportional to particle multiplicity crossing
the i-MCP and, in turn, to the energy deposited in the $3X_0$
absorber. This relationship defines the amplitude to energy
calibration of the MC simulation. After calibration, events are
generated scanning several energies from 10 to 300 GeV. The energy 
deposited in the preshower in each event is estimated from the
amplitude and added to the energy deposited in a calorimeter block
downstream of the preshower, with ideal resolution. With this
procedure, the resulting energy sum is only smeared by the amplitude
response spread due to the preshower, and is therefore suited to
predict the resolution term that this preshower would add to the
calorimeter. According to simulation, we find this contribution to be
smaller than 5\%/$\sqrt{E \rm{(GeV)}}$, which is definitely a small
contribution for most electromagnetic calorimeters. 

\section{Summary and outlook}
\label{conclusion} 

We report on the response of microchannel plates (MCPs) to single
relativistic particles and to electromagnetic showers. Several
prototypes of photodetectors with the amplification stage based on
MCPs were exposed to cosmic rays and to 491~MeV electrons at the
INFN-LNF Beam-Test Facility. The MCPs were used as secondary emission
detectors, by applying a retarding bias to the photocathode. In this
configuration, time resolutions of about 50~ps with cosmic muons, and
detection efficiencies to single relativistic particles of order 50\%
are obtained. Measurement with electromagnetic showers sampled at
different depths shows that the MCPs efficiency has a simple binomial
dependence on the multiplicity of particles in the shower. Starting
from the interpretation of the results, lines of investigations to
further enhance the response of detectors relying on secondary
emission from the MCP surface are suggested. While these
investigations are being pursued, we show that present results make
this detection technique worth considering for application in the
precision timing of high energy photons and charged particles, to aid
in event reconstruction at high luminosity colliders.

\section*{Acknowledgements}
We warmly thank M. Barnyakov for providing us with and for valuable
information on the Ekran FEP photomultipliers used in this study. We
gratefully acknowledge the skilful help and continuous support of
R.~Ber\-to\-ni, R.~Mazza, M.~Nuc\-ce\-tel\-li and F.~Pel\-le\-gri\-no
for the preparation of the experimental setup. We are indebted to
B.~Buo\-no\-mo,  L.~Fog\-get\-ta and P.~Va\-len\-te for their help
with the setup of the beam facility. We also thank our students
A.~Beschi, S.~Bologna, M.~Defranchis and M.~Gregori for valuable
contributions. This research program is supported by INFN CSN5.  

\bibliography{imcp_btf}{}
\bibliographystyle{unsrt}
\end{document}